# Visual Modulation of Human Responses to Support Surface Translation

Mustafa Emre Akçay[1], Vittorio Lippi[2]* and Thomas Mergner[2]*

[1] Department of Mechatronics, Engineering, Kocaeli University, Kocaeli, Turkey, [2] Neurological University Clinics, University of Freiburg, Freiburg, Germany

Vision is known to improve human postural responses to external perturbations. This study investigates the role of vision for the responses to continuous pseudorandom support surface translations in the body sagittal plane in three visual conditions: with the eyes closed (EC), in stroboscopic illumination (EO/SI; only visual position information) and with eyes open in continuous illumination (EO/CI; position and velocity information) with the room as static visual scene (or the interior of a moving cabin, in some of the trials). In the frequency spectrum of the translation stimulus we distinguished on the basis of the response patterns between a low-frequency, mid-frequency, and high-frequency range (LFR: 0.0165-0.14 Hz; MFR: 0.15–0.57 Hz; HFR: 0.58–2.46 Hz). With EC, subjects' mean sway response gain was very low in the LFR. On average it increased with EO/SI (although not to a significant degree $p = 0.078$) and more so with EO/CI ($p < 10^{-6}$). In contrast, the average gain in the MFR decreased from EC to EO/SI (although not to a significant degree, $p = 0.548$) and further to EO/CI ($p = 0.0002$). In the HFR, all three visual conditions produced, similarly, high gain levels. A single inverted pendulum (SIP) model controlling center of mass (COM) balancing about the ankle joints formally described the EC response as being strongly shaped by a resonance phenomenon arising primarily from the control's proprioceptive feedback loop. The effect of adding visual information in these simulations lies in a reduction of the resonance, similar as in the experiments. Extending the model to a double inverted pendulum (DIP) suggested in addition a biomechanical damping effective from trunk sway in the hip joints on the resonance.

Keywords: human posture control, vision, support surface translation, balance control, modeling

*Correspondence:
Vittorio Lippi
vittorio_lippi@hotmail.com
Thomas Mergner
thomas.mergner@uniklinik-freiburg.de

Citation:
Akçay ME., Lippi V and Mergner T (2021) Visual Modulation of Human Responses to Support Surface Translation. Front. Hum. Neurosci. 15:615200. doi: 10.3389/fnhum.2021.615200

## INTRODUCTION

Humans use mainly vestibular, proprioceptive, and visual cues for their balancing of upright stance (Horak and Macpherson, 1996) and adjust the use of these cues depending on the external disturbances and environmental conditions in terms of a "sensory reweighting" (Nashner and Berthoz, 1978). This study is focused on the contribution of vision to the control of standing balance during support surface translation in the body sagittal plane. Generally, early observations already indicated that a stationary visual space reference reduces spontaneous and externally evoked body sway (Romberg, 1846; Edwards, 1946; Paulus et al., 1984) while the motion of a visual scene in turn may evoke sway responses (Lee and Lishman, 1975; Berthoz et al., 1979).

Studies emphasizing the importance of vision for balancing control often relied on tests of patients with a degraded vestibular or somatosensory sense (Romberg, 1846; Travis, 1945; Edwards, 1946; Paulus et al., 1984). Also, the motion of real or virtual visual scenes has been used to evoke postural sway in normal subjects to analyze parameters relevant for evoking the visual effect (Lee and Lishman, 1975; Lestienne et al., 1977; Berthoz et al., 1979; Dichgans and Brandt, 1979; Soechting and Berthoz, 1979; Bronstein, 1986; Van Asten et al., 1988). Peterka and Benolken (1995) isolated the effect of visual cues from vestibular and proprioceptive effects by presenting tilt of visual scenes in the BSRP[1] condition to vestibular loss (VL) and vestibular-able (VA) subjects. Interestingly, to improve the balancing capability of the VA subjects' the visual tilt magnitude had to exceed the value of their vestibular threshold.

In earlier work of Berthoz et al., 1979, Buchanan and Horak (1999, 2001), and Corna et al., 1999, the authors used sinusoidal support surface translations when studying the stabilizing effect of vision on translation-evoked body sway (by comparing in an illuminated stationary laboratory the balancing performance with eyes closed, EC, to that with eyes open, EO). With sinusoidal support surface translations at low frequency (0.2 Hz), subjects were "riding the platform," meaning that they were maintaining the body erect and upright, but involved the hip joints in the body stabilization at higher stimulus frequencies. Pointing out that the periodicity of the sinusoidal stimuli in these studies may have allowed subjects to involve prediction in their disturbance compensation (see Dietz et al., 1993) and Jilk et al. (2014) used the pseudorandom ternary sequence, PRTS, stimulus of Peterka (2002; see **Figure 1**), since this contains in overlay a broad spectrum of frequencies and typically is considered by subjects as unpredictable.

Jilk et al. (2014) found with translation stimuli upon switching from EC to EO a shift from a SIP to a DIP (double inverted pendulum) mode of balancing, and this, similarly, when increasing translation amplitude and/or frequency. They concluded that, irrespectively, of considerable inter-individual variations in the use of the ankle versus the hip contribution, the relevant control variable was the balancing of the body COM. However, they left open in which way the hip supports the balancing of the body COM in the ankle joints. The present study elaborates on this issue, accompanying our experimental with a modeling approach.

Formal descriptions of the involved sensorimotor mechanisms are still a topic of ongoing research. The present study investigates the postural response to support surface translation in the body sagittal plane with the aim to formally describe the effects of vision on the response using different visual viewing conditions. For comparison, for the formal description of tilt responses, Peterka (2002) designed the Independent Channel, IC, model. This formal description covers in addition to changes in tilt amplitude also the sensor availability for vision (eyes closed, EC, versus eyes open, EO), and the body support condition (e.g., firm versus body sway referenced platform, BSRP[1]). By this, the IC model allows replicating sway responses observed in human tilt experiments using a simple sensory feedback control scheme that linearly combines weighted sensory cues.

Such experimental data can alternatively be described using the disturbance estimation and compensation (DEC) model (Maurer et al., 2006; Mergner et al., 2009). The DEC model is more complex than the IC model in that it feeds back disturbance estimates for compensation,[2] and in that it contains non-linear elements (detection thresholds). But it is also more general in that it covers other external perturbations such as support surface translation (Mergner, 2010) and in that it can control several degrees of freedom of the body at the same time. Implemented into humanoid robots as proof of principle, it produced coordinated ankle and, additionally, hip responses to support surface tilt in a 2 DoF robot (Hettich et al., 2014) and coordination across the sagittal and frontal body planes in a 14 DoF robot (Lippi and Mergner, 2017).

Generally, model description can help to better understand biological functions. Assländer et al. (2015) used the DEC model when addressing the question whether visual position and velocity information serve different functions in the postural control of tilt responses. The authors compared tilt-evoked sway responses in the presence of visual velocity and position information (continuous illumination) with only visual position information available (stroboscopic illumination; see Paulus et al., 1984; Amblard et al., 1985; Assländer et al., 2013). This allowed them to associated the visual velocity signal with an improved compensation of the external stimulus (tilt) and the visual position signal with the compensation of the gravity effect arising with body lean. These visual improvements took effect by an increase in gain and a lowering of a detection threshold in the disturbance estimation and compensation mechanisms. Comparable work on the postural responses to support surface translation is still missing to date.

Conceptually, visual improvements of postural control have in the past often been attributed to a reduction of sensory noise in the system, presumed to stem foremost from the vestibular signal. To formally describe the phenomenon, multi-sensory integration methods such as Kalman filtering or the Bayesian estimation theory have been used (Van der Kooij et al., 1999, 2001; Dokka et al., 2010). The noise of vestibular cues has been estimated to be about ten times larger than that of proprioceptive cues (van der Kooij and Peterka, 2011). In ego-motion perception, observed detection thresholds were considered useful in that they prevent noise from producing self-motion illusions at rest (Mergner et al., 1991). Similarly, in the DEC model they are thought to shield the disturbance estimates at least partially from noise during support tilt (Mergner, 2010).

In the present context of support surface translation stimuli, however, vestibular noise appears to play no major role. As emphasized in the study of Lippi et al. (2020), the physical and sensory impacts evoked by support surface translation differ from those evoked by tilt. In particular, vestibular and proprioceptive signals maintain during translation on level and firm support a fixed relation to each other, so that subjects may choose to use for balancing the less noisy proprioceptive signal of body lean with respect to the support instead of the noisy vestibular signal of body lean in space. Therefore, we considered in the present context of support translation the vestibular noise aspects to be of little relevance. In order to corroborate this view, we implemented the DEC concept not only as a simulation model, but tested this model also in a 2 DoF humanoid robot for robustness against noise and technical inaccuracies.

---

[1] During evoked or spontaneous body sway with BSRP, the sway signal is recorded and used to tilt the support with the body such as to minimize ankle joint excursion and proprioceptive stimulation (Nashner and Berthoz, 1978). When subjects with severe loss of vestibular function then close their eyes, they tend to fall, unlike subjects with normal vestibular function.

[2] Their functionality is here, noticeably, equivalent to compensatory feed-forward mechanisms. This type of mechanism is denoted in German textbooks of control theory as "Störgrößen-Aufschaltung."

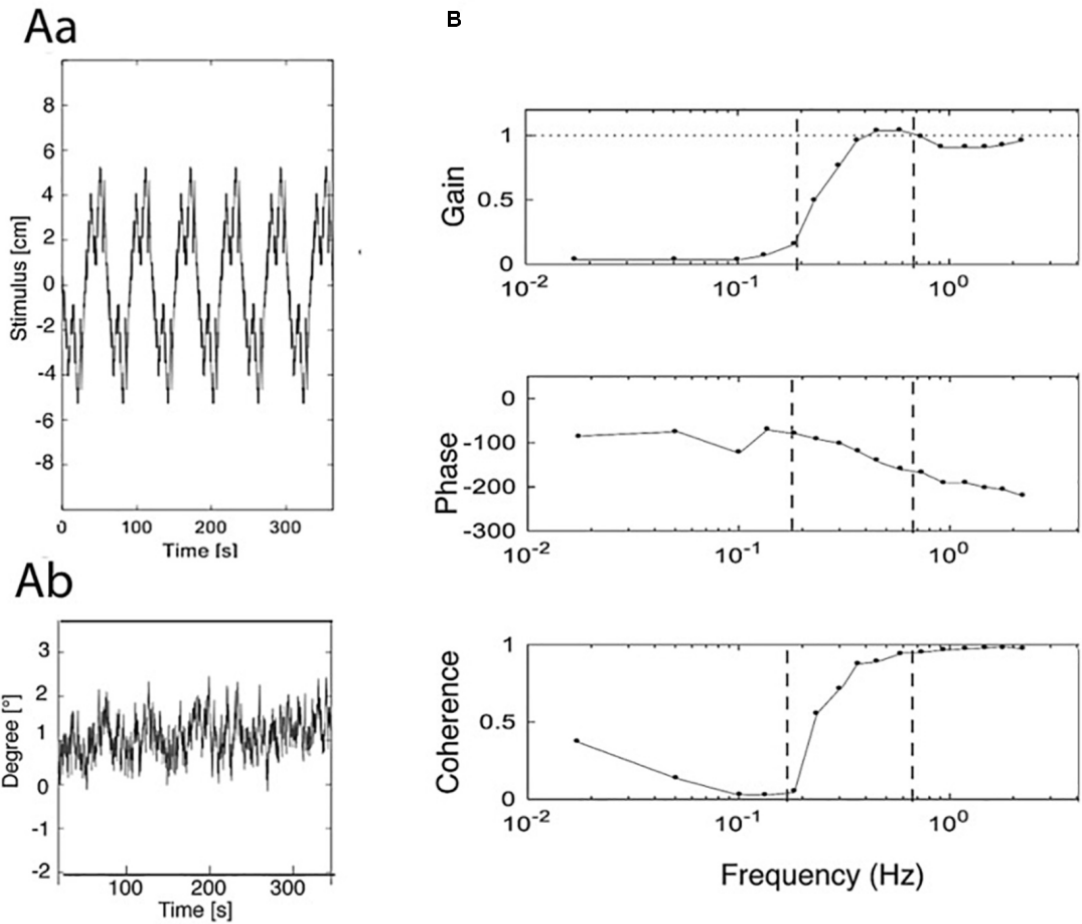

**FIGURE 1 |** Examples of stimulus, original response, and frequency response functions (FRFs). **(Aa)** PRTS translation stimulus (pp. 9.6 cm). **(Ab)** Evoked sway response of body COM around ankle joints (eyes closed). **(B)** Derived frequency response functions (FRFs) of gain, phase, and coherence. Vertical dashed lines separate low, middle, and high-frequency ranges of the stimulus; compare **Figure 2**). Gain represents the amplitude ratio between sway response amplitude and translation stimulus amplitude. Unity gain indicates a response of 1 sway amplitude per 1 cm of support translation and zero gain the absence of any evoked sway. Phase reflects the temporal relation between the translation stimulus and the sway response (0 indicating exactly in-phase response and –180 a lean counter to platform translation). Coherence is a measure of the signal to noise ratio of the stimulus-evoked sway (the coherence for a linear ideal system with no noise would be unity over the whole spectrum). To get an impression of the stimulus, see https://youtu.be/BUlcKQ67JCk.

The quality of the visual stimulus used in such studies is, however, often of relevance. In the experiments of Mergner et al. (2005), the tilt of a virtual reality room was, even after full perceptual immersion into vection (self-motion perception), less effective in producing body sway than the tilt of a moveable physical room (e.g., represented by an internally illuminated cabin). Yet, using the latter was less effective for improving balance control compared to the illuminated laboratory room – which underlines an important contribution of cognition (Blümle et al., 2006). As a consequence for the present study, we used in the main experiments of the present study the scene of the laboratory room as visual space reference.

Taken together, the present study aims to formally describe the role of vision for human postural responses to support surface translations such that model and robot simulation results approximately match the experimental results. With the focus on the role of vision for the translation response, we distinguished between the effects of visual position versus velocity plus position cues by using stroboscopic (flashed) versus normal (continuous) illumination. Using for support surface translation stimulus a continuous pseudorandom ternary sequences (PRTS) translation profile allowed us to cover a broad band of frequencies of the postural responses. Our modeling of the responses suggests that these were shaped in terms of an amplitude enhancement to a large extent by a resonance of the proprioceptive feedback loop controlling the ankle joints, and that vision of a stationary scene is damping this resonance together with a biomechanical effect from hip bending.

## MATERIALS AND METHODS

The experiments were performed in the posture control laboratory of the Neurological Clinic, University of Freiburg, Germany. The study was approved by the Ethics Committee of the Freiburg Clinics and was conducted in accordance with the 1964 Declaration of Helsinki in its latest revision.

### Subjects

Seven subjects (3 female, 4 male; 24.4 5.1 years of age) participated in the study after giving written informed consent. Subjects' mean mass and height were *76.8* 7.5 kg and 1.73 0.08 m, respectively. All subjects were without a history of balance impairment and epilepsy (the latter was explicitly requested since the stroboscopic illumination used has the potential to evoke epileptic seizures). Each subject performed the experiments twice (two trials recorded and used to produce the average responses for each condition). The sensitivity of the performed tests (ANOVA) with the given number of subjects was analyzed. Setting the power to b = 0.80 and the significance level a = 0.005 the minimum detectable effect size was 81% of the common standard deviation within the groups.

### Experimental Setup

Subjects stood on a motion platform holding in each hand a handle of a safety rope, which gave no support or spatial orientation cues with flexed arms during trials but supported the body when the arms were stretched (compare Lippi et al., 2020). Body kinematics were measured using active markers of an optoelectronic motion capturing system (Optotrak 3020; Waterloo, ON, Canada), attached to subjects' hips and shoulders and to the platform (details in Assländer et al., 2015). A PC with custom-made programs was used to generate the support translation stimuli in the body-sagittal plane. Another PC was used to record the stimuli and the body sway from the marker positions at a sampling frequency of 100 Hz using software written in LabView (National Instruments; Austin, TX, United States).

The experiments were performed with the eyes closed (EC) or open (EO). With eyes open, we used for the visual scene two conditions. In one set up condition, the visual scene was represented by the interior of the illuminated laboratory room (assuming that subjects experience it as part of the building and thus as a visual representation of physical space). In this situation, the visual stimulus arose from the motion of the body on its translating support with respect to the lit interior of the stationary walls of the laboratory room while looking straight ahead with a distance of approximately 3 m to the front and to each side wall (with minor changes from the translation-evoked sway of the body; the translation stimulus amplitude always amounted to pp. 9.6 cm, see below). In the second set up, the visual scene consisted of the interior of a lightweight cabin (2.00 m height 1.1 m wide and 1.0 m length in front of the subject, with inside illuminated wall patterns). Using the cabin, two visual motion conditions were used (translation amplitudes as before pp. 9.6 cm): (a) Translation of the cabin in space, with the subject standing on stationary external support and viewing translation of the cabin's interior visual scene ("visual only stimulus." (b) *En-bloc* translation of cabin and platform (with the subjects standing on the platform, they visually perceived the evoked body sway, but not the body support translation). Overall, seven experimental stimulus conditions were used (**Table 1**). Stroboscopic illumination (4 Hz; flash duration of 5 ms) was applied in the trials with reduced visual velocity cues, EO/SI (using the method and equipment of Assländer et al., 2013, 2015). The translation stimulus was always applied in the body-sagittal plane. It lasted 363 s and contained six repetitions of a pseudo-random ternary sequence motion (PRTS; see Peterka, 2002) with 242 states and a state duration of 250 ms.

The amplitude spectra of support and/or scene displacement, velocity, and acceleration of the stimulus are given in **Figure 2**. Based on the spectral characteristics of the stimulus (**Figure 2**) we distinguished between three frequency ranges: Low-frequency range (LFR: 0.0165–0.14 Hz), mid-frequency range (MFR: 0.15– 0.57 Hz), and high-frequency range (HFR: 0.58–2.46 Hz). Denoting that the physiological relevant stimulus parameter is here the acceleration dependent inertial force acting on the body, this division equally separates the respective low, mid, and high acceleration ranges, respectively.

**TABLE 1** | The visual and translation stimulus conditions tested (EO, eyes open; EC, eyes closed; onlyCab, only cabin with interior scene is translated, body support is stationary; Cab&Body, en-bloc translation of cabin with scene and body support; details in text).

| Stim. | Eyes | Visual scene | Illumination | Body support |
| --- | --- | --- | --- | --- |
| 1. EC | Closed | – | – | Translation |
| 2. EO/SI | Open | Stationary | Stroboscopic | Translation |
| 3. EO/CI | Open | Stationary | Continuous | Translation |
| 4. onlyCab/SI | Open | Moving cabin – body stat. | Stroboscopic | Stationary |
| 5. onlyCab/CI | Open | Moving cabin – body stat. | Continuous | Stationary |
| 6. Cab&Body/SI | Open | Moving cabin and body | Stroboscopic | Translation |
| 7. Cab&Body/CI | Open | Moving cabin and body | Continuous | Translation |

Each condition was tested twice per subject.

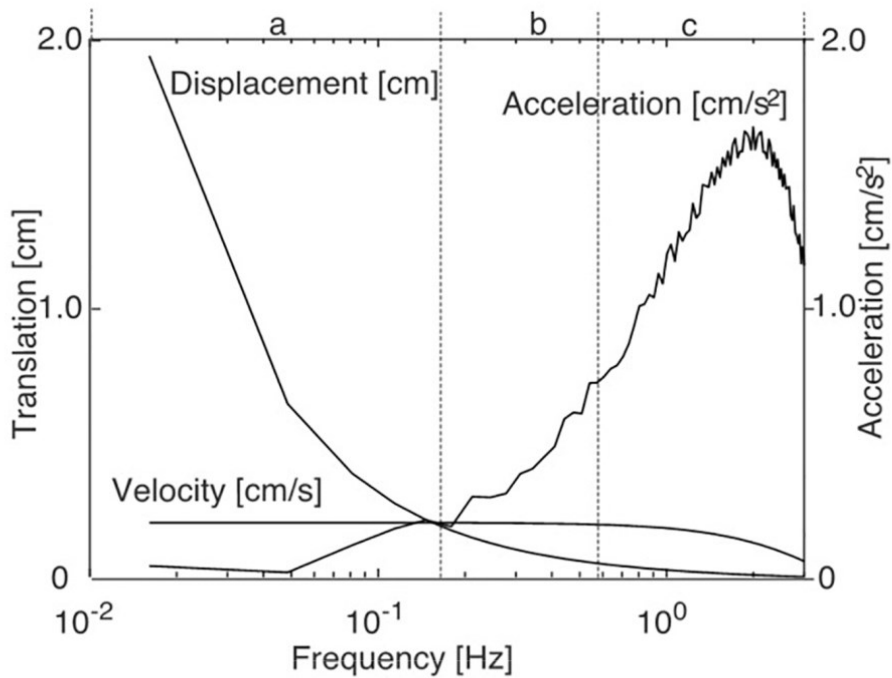

**FIGURE 2** | Spectral characteristics of the PRTS (pseudo-random ternary sequence) translation stimulus. a, b, and c give the low, mid, and high frequency (and acceleration) ranges referred to in the text and the Results figures.

## Procedures

In pilot experiments, we familiarized subjects with the experimental setup and the stimuli. In these experiments, we observed that the sway response to the translation stimulus involved a considerable upper body (or HAT, for head, arms, and trunk) sway on the hip joints (see translation amplitude, which amounted here to pp. 9.6 cm, but is also in line with previous findings where we used considerably smaller amplitudes; Lippi et al., 2020). We considered the HAT sway foremost a physiological biomechanical effect, but rejected to suppress it by using a backboard since this tended to interfere in pilot experiments with the normal sway response.

Subjects were instructed to stand upright and comfortable and to look straight ahead or to close their eyes, depending on the stimulus condition. During trials, subjects wore headphones and listened to audiobooks to avoid auditory spatial orientation cues and to distract attention away from the balancing, to let this happen automatically. In each experimental session, 4–6 trials were tested, separated by 1–3 min breaks, during which subjects were instructed to perform relaxing arm and leg movements. Intervals between trials were between 5 and 20 min, and 6–8 experimental sessions were distributed over several days. Translation amplitude here always amounted to
pp. 9.6 cm (a stimulus magnitude that subjects considered as comfortable).

## Data Analyses

All analysis steps orient on the methods proposed by Peterka (2002).

### Time-Domain

Recorded sway kinematics were exported to Matlab (The Mathworks, Natick, MA, United States) for analysis. Sway kinematics of the leg and the HAT (head, arms, and trunk) segment were obtained from the hip and shoulder marker translations and the manually measured marker heights using trigonometric functions. Angular sway of the body center of mass (COM; excluding the feet) was obtained from these measures and the anthropometric data of Winter (2009). The sway of the HAT segment in the hip joints as well as of the leg segment and the COM around the ankle joints (in ) were related in the following analysis steps to the linear horizontal excursions of the support base and/or of the visual scene (body-sagittal translations, in cm).

### Frequency Domain

The frequency-domain representation of the variables (stimulus sequence; leg, HAT, and COM sway) was obtained using the fast Fourier transform implemented in the Matlab function 'fft'. After averaging the spectra across stimulus cycles, frequency response functions (FRFs) were calculated dividing output (leg, HAT, or COM sway spectrum) by input (translation stimulus spectrum) for each frequency point and expressing the result in terms of gain (absolute value) and phase (inverse tangent of FRF; Pintelon and Schoukens, 2004). Coherence functions were calculated dividing the squared absolute value of the averaged cross power spectrum by the product of the averaged input and output power spectra. The PRTS stimulus has the property that only odd frequency points of the discrete spectrum with the fundamental harmonic at 0.0165 Hz have stimulus energy (Davies, 1970) while even harmonics do not contain stimulus power and are not included in the analysis. For the results presented in **Figures 1, 3** below, we averaged gain, phase, and coherence curves across frequency in approximately logarithmic distanced bins (compare Peterka, 2002). The sway responses of subjects are presented in the following as frequency response functions (FRFs) for gain, phase, and coherence. One way Anova statistics was performed for the averages of the FRF gain to test for differences between the LFR and MFR ranges. The average of the gains over the frequency range of interests was used as variable in the comparisons instead of performing a multivariate analysis considering the gains at each frequency. This was motivated by the idea of defining a single variable expressing the concept of "having a different gain" over a frequency range. We acknowledge that this approach may be conservative, in the sense that differences over a specific frequency can be overlooked once the gains are averaged.

## Simulations

Model simulations were performed with two models: (1) A *single inverted pendulum (SIP) model* (lower half of **Figure 4** activated, while the upper part is inactivated; a detailed description of the active part is given in the **Supplementary Appendix**). (2) A DIP model (both lower and upper parts in **Figure 4** activated). Both models were used to formally describe the translation evoked sway of the body COM in the ankle joints, and the DIP model to describe in addition the effects of passive and reflexive hip bending. The initial condition for the simulations were the upright body pose. The DIP model was derived from the work of Hettich et al. (2014). It was previously used in our laboratory to describe ankle and hip responses to support surface tilts. For the relative motion between the upper segment (HAT, for head-arms-trunk) and the lower body we used here simply high proprioceptive and passive hip stiffness to the effect that the HAT segment was maintained approximately aligned with the lower body with slow stimuli. The SIP model was additionally implemented into a humanoid robot for a 'real-world robustness' test (i.e., for testing the balancing performance in the presence of technical noise and inaccuracies) using the same motion platform as for the human subjects.

For the robot simulations, we used the humanoid robot Posturob II which operated with the DEC control (antropometrics similar to humans: height 1.72 m, weight 56 kg; see Hettich et al., 2014). In short, the robot actuates hip and ankle joints with pneumatic "muscles" using torque control. It uses technical analogs for joint proprioception (potentiometers), an artificial vestibular system (Mergner et al., 2009), and ankle torque sensors (force transducers). As a substitute for a visual sensor, we used the sign-reversed signal of body sway in space from the Optotrak system. The sensory signals were digitized with an AD card and implemented on a PC running as Real-Time Windows Target (The MathWorks Inc., Natick, MA, United States). The sensory signals were used as inputs for the control model. The model output was the torque command for the ankle joints, after converting it into an analog signal using a DA card. The initial condition for all simulations were the upright body pose.

## RESULTS

An overview over stimulus, data collection, and analysis is given in **Figure 1**. It shows the three processing steps from stimulus to response, depicting the support displacement stimulus (six repetitions of the pp. 9.6 cm PRTS stimulus; Aa), the evoked COM sway in the body sagittal plane around the ankle joints in the EC condition (Ab), and the processed FRFs of gain, phase, and coherence (B).

The FRF results obtained for the seven visual stimulus conditions listed in **Table 1** are depicted in **Figures 3A–G**. Panels A-C give the body COM sway results for the support translation in the laboratory room. Panel A shows that the stimulus evoked with eyes closed (EC) in the low-frequency range (LFR) almost no sway (gain 0.1; reflecting almost upright body COM). In the mid-frequency range (MFR), gain shows an abrupt increase with a peak slightly exceeding unity gain (note also increase in coherence). This gain increase was statistically significant ($p < 0.001$). In the high-frequency range (HFR), gain levels off to a value slightly below unity. The phase passes in the transition from the mid to high-frequency range a lag of 180 (corresponding to a lean of the body COM counter to the stimulus).

With stroboscopic illumination of the laboratory (EO/SI), providing visual position information on self-motion, the patterns of the phase and the gain FRFs (**Figure 3B**) basically resembled those obtained without visual information, as shown in **Figure 3A** (the difference was not statistically significant: $p = 0.078$), but gain is slightly increased in the low-frequency range and reduced in the mid-frequency range. These changes became enhanced when allowing additionally for visual velocity information in continuous illumination (panel C; EO/CI). Noticeably, however, in the high-frequency range, gain reached a similar value close to unity with all three illumination conditions (note arrows at the endpoint of the gain FRFs in the three panels).

As described in Introduction, viewing a moving visual scene may evoke body sway responses on the background of self-motion illusions. Here, the cabin was translated with the PRTS stimulus profile while the subjects stood on the stationary body support viewing the moving cabin's interior visual scene. This visual stimulus-evoked only very small sway responses mainly in the low-frequency range, this with stroboscopic illumination (EO/SI; **Figure 3D**) and only slightly more so with visual velocity and position cues (EO/CI; **Figure 3E**; note also low coherence curves). On retrospective request, all subjects reported for these stimuli a perception of scene motion and none of them a self-motion perception. Depicted in **Figures 3D,E** are in addition the results of our model simulations with our SIP model (dotted gain curves in **Figures 3D,E**). The gain of these simulated responses is clearly larger than that actually obtained (full lines). We attribute this difference to unmodeled human high-level cognitive mechanisms suppressing the occurrence of vection in our test conditions with the jerky PRTS translation stimulus. Thus, using here isolated visual scene motion as stimulus did not contribute substantially to a better understanding of the role of vision for the balancing of support translation.

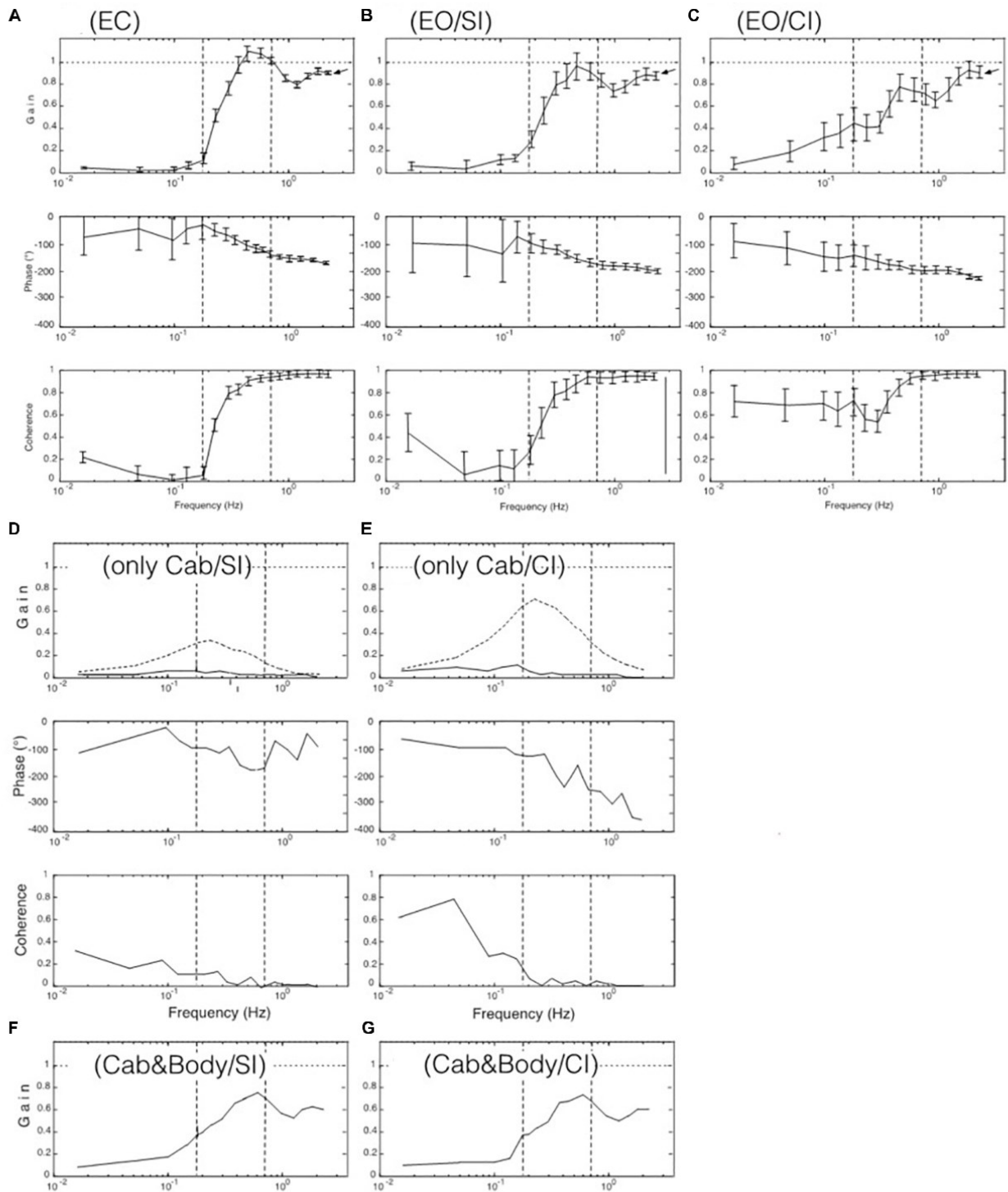

**FIGURE 3** | Results from the seven experiments reflecting the seven tested viewing conditions listed in **Table 1**. Shown are gain, phase, and coherence (in **A–C** with standard error values) in terms of gain, phase, and coherence over frequency. Dotted curves in panels **(D,E)** demonstrate discrepancy of obtained versus modeled data (see text). Curves in **(F,G)** represent "distorted" responses obtained with the cabin fixed on the translating body support.

The results obtained in the experiments with the cabin fixed on the translating body support (Cabin&Body) are shown in **Figures 3F,G**. This stimulus was considered as particular under the aspect that vision of the relative body-to-scene translation was missing while that of the evoked body sway was maintained (like in an accelerating lit underground cabin). The responses deviated substantially from the results in **Figures 3B,C** in that a considerable difference across the SI and CI conditions was missing and the overall gain was reduced. We take this as evidence that subjects tended to shape their translation responses in terms of a damping effect not only from vision, but also from hip joints motion and present biomechanical evidence and corresponding model simulations in **Supplementary Appendix B**.

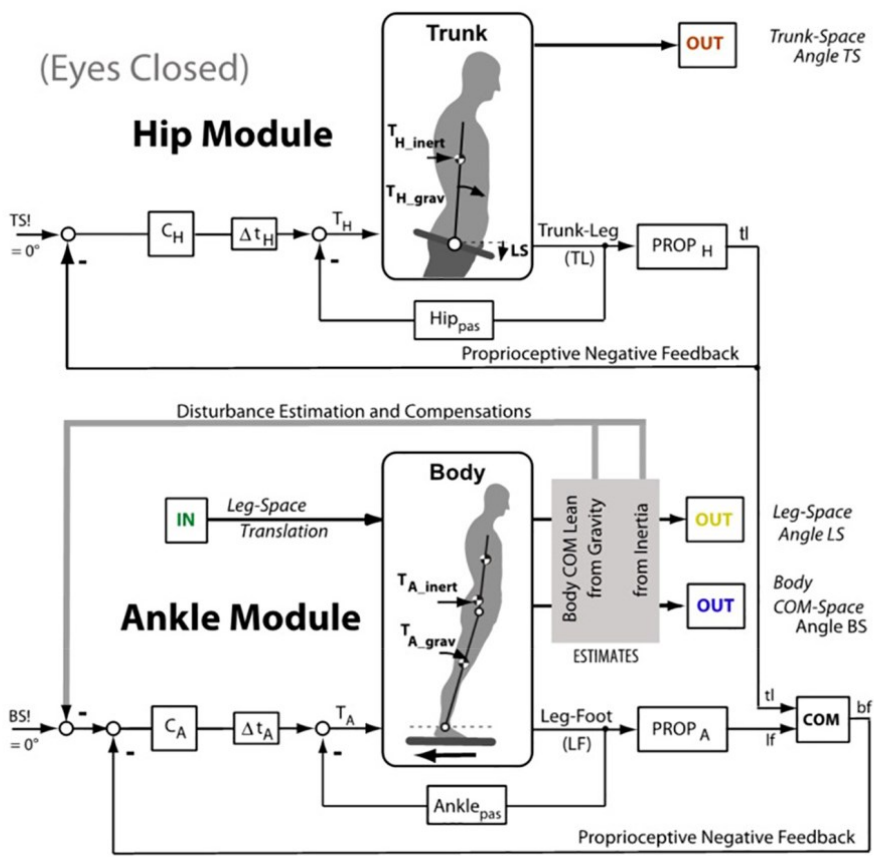

**FIGURE 4** | Inverted pendulum model (simplified version of the model used by Hettich et al. (2014). The lower half (Ankle Module) shows the control of the body center of mass (COM) with negative feedback from proprioception and passive tissue (pas). The input is desired body COM orientation in space (BS!; 0, upright). The controller $C_A$ (proportional-derivative) provides after the lumped time delay $\Delta t$ the ankle torque $T_A$ that stabilizes at rest the upright body pose through the feedbacks from ankle proprioception ($PROP_A$) and passively from connective tissue ($Ankle_{pas}$). During support surface translations, deviations from the earth vertial body pose by the translation evoked inertial and gravitational ankle torque ($T_{A\,inert}$; $T_{A\,grav}$) are compensated for by a Disturbance Estimation and Compensation (DEC) mechanisms (gray box; details in **Supplementary Appendix A**). The role of the upper body stabilization on the hip joints for whole body COM stabilization follows comparabale principles and is adding to the sway of the body COM. In the context of the present experiments, we considered two modeling scenarios, one where the gain of the hip module was so much enhanced that essentially a SIP sway in the ankle joints resulted, and the other where a DIP scenario with additional sway in the hip joints resulted. In the latter case, a damping of the body sway responses from support translation in the ankle joints resulted, adding to sway damping resulting from visual input (details in **Supplementary Appendix B**). The figure emphasises that the balancing involves both, the ankle joint control and the hip joint control.

## Model Simulation Results

Using the DEC model shown in **Figure 4** (see also **Supplementary Appendix A**), simulations were performed comparing, by visual inspection, EC responses with EO/SI and EO/CI responses (initial condition was upright body pose). This approach assumes that a multisensory signal "sway of the body COM" is the control variable for the balancing (compare Jilk et al., 2014). The visual input in these simulations provided information on body COM motion with respect to inertial space (mimicking vision of the illuminated stationary laboratory and its use as space reference). The results are shown in **Figures 5A–C**. For the results shown in panel A, no visual input was used. For the results in panel B, visual position input was added and used for compensation of the gravity effect with body lean. For the results in panel C, furthermore visual velocity input was added to improve the gravity (body lean) estimate as well as the vestibular estimate of support translational acceleration (see Appenix A). With appropriate gain adjustment of these visual contributions, the simulation results reproduced the main response features obtained in the human experiments (**Figures 5A–C**).

Considering the relative contributions of the DEC for compensation of the gravity effect versus that for the acceleration force, we found that the contribution of the latter was relatively small. Generally, its gain requires a downgrading to avoid control instability in face of the second derivative term (acceleration) in the sensory feedback (see model in **Supplementary Appendix 1**).

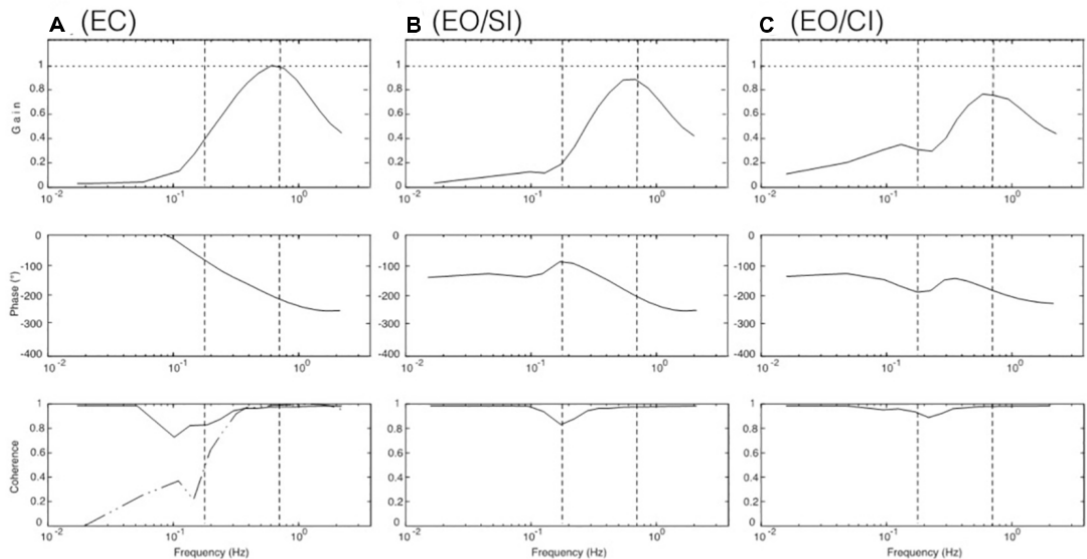

**FIGURE 5** | Model simulation results for COM balancing without vision (**A**; EC), with visual position information available (**B**; EO/SI), and with both visual position and velocity information available (**C**; EO/CI). Note that these results reproduce the main response features obtained in the human experiments (compare **Figures 3A–C**). The dash-dot coherence curve in (**A**) shows result of adding low-frequency noise to the control. It suggests that noise may explain to a large extent the difference in coherence seen between the simulations and the human data in the low-frequency range.

## Robot Simulation Results

The robot simulations were performed using the humanoid robot Posturob II (see section "Materials and Methods") in the human laboratory using the same stimuli and analyses as described for the human subjects. The results for the body COM sway obtained in the absence ("EC") and the presences ("EO/CI") of visual information are given in **Figures 6A,B**. Note that they show very similar response characteristics as those of the human subjects (compare **Figures 3A,C**) and the model simulations (**Figures 5A,C**).

## DISCUSSION

This study's aim is to formally describe how humans adjust their sensorimotor control when they use their joint proprioception combined with visual cues for balancing of erect stance during horizontal support surface translations in the body sagittal plane. As a means to shed light on this control from different perspectives, we used the three viewing conditions EC, EO/SI, and EO/CI. The DEC model (**Figure 4** and **Supplementary Appendix A**) was used to formally describe the experimental data. This description will be used in the following to formulate a working hypothesis on the role of the visual cues in the human balancing of support surface translation.

In the following, we address: (A) A concept on how to interpret the sequence of a reduced sway in the LFR followed by a pronounced gain peak in the MFR in the EC condition, with both effects showing a diminished manifestation in the EO conditions. Our hypothesis is that the gain peak represents a *resonance phenomenon*, which is damped when visual cues became available. This implies (B) the question in which way the visual cues were reducing the resonance tendency and what the role of involving hip motion in this respect may be. (C) Finally, we focus on more general considerations such as the comparisons with earlier work that used sinusoidal and thus predictable (and theoretically also resonating) translation stimuli and (D) on biomechanical aspects.

## Resonance Hypothesis

Our hypothesis is that the gain peak in the mid-frequency range in the EC condition (**Figure 3A**) results from a resonance phenomenon in the sensory feedback control of the ankle control. In mechanics, resonance is the tendency of a system to respond with a greater amplitude when the frequency of its oscillations matches the system's natural frequency, which eventually may endanger control stability. In the model shown in **Figure 4**, resonance relates primarily to the proprioceptive negative feedback loop in the ankle joint control depending on its intrinsic time delay and the chosen gain value. Our hypothesis is inspired by the earlier work of Peterka and Loughlin (2004) who showed that human sway responses, evoked by producing abrupt discrepancies between required and actual sensory feedback gain, tend to show transient resonance oscillations. When the authors set in experiment-driven simulations the loop gain of the control with sensory feedback to a low value, the system showed a tendency for oscillations at about 0.1 Hz, and with high gain for oscillations at about 1 Hz (which approximately corresponds to the gain peaks seen in the results of our study; see **Figures 3A–C** and also the frequency response functions below in **Figures 7A–C**). We elaborated on this hypothesis by repeating the model simulations with several modifications.

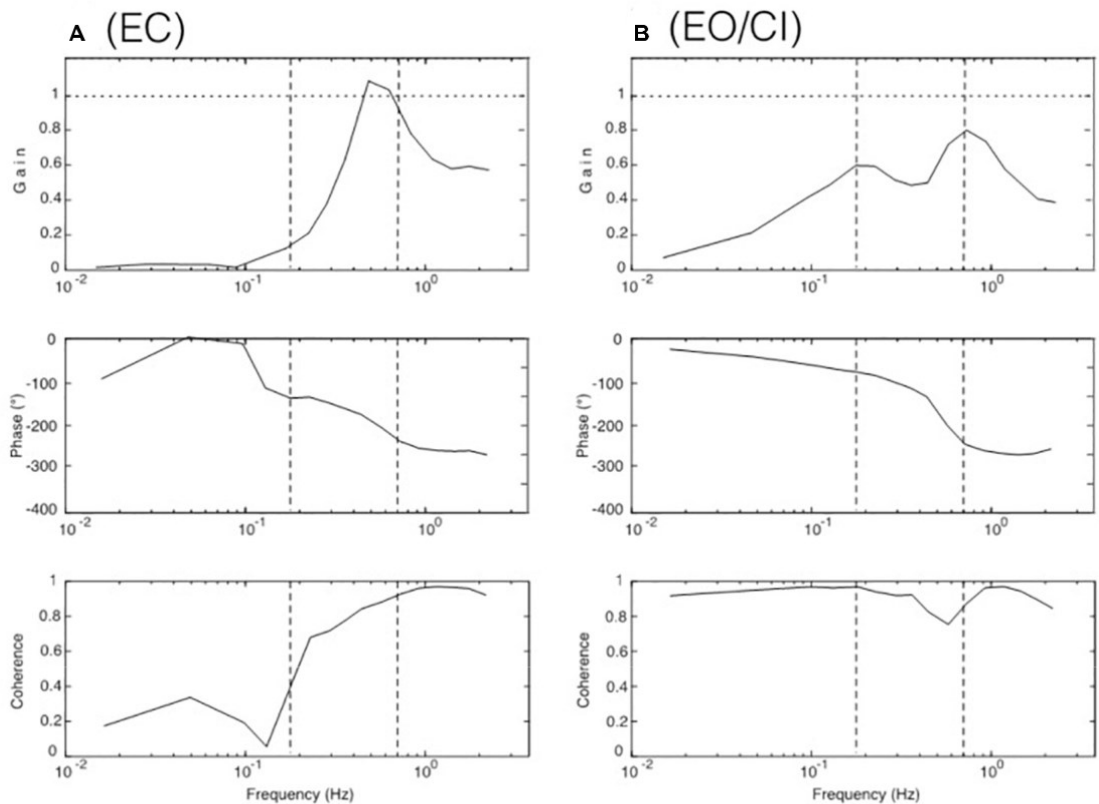

**FIGURE 6** | Robot simulation results for COM balancing without vision (**A**, EC for "eyes closes") and with visual input corresponding to the continuous illumination condition (**B**, EO/CI).

In one set of simulations, we focused on changes of the loop gain (i.e., of the combined gain of the passive and the proprioceptive feedback stiffness; see model in **Figure 4**). In simulations with the corresponding SIP model we obtained results consistent with the notion of a resonance tendency, demonstrating with the translation stimulus a resonance peak at about 0.7–0.8 Hz with high loop gain and, as evidence for a second resonance with low loop gain, a smaller peak at 0.16 Hz (**Figure 7A**). Increasing/decreasing in these simulations instead of the gain the time delay had, in contrast, mainly an increasing/decreasing effect on peak height (**Figure 7B**), while increasing selectively the gain of the passive stiffness shifted the peak mainly across frequencies (**Figure 7C**).

Conceiving that the resonance may be initiated and maintained by the to-and-fro of the translation stimulus, we extended our simulations and injected white noise into the control input, after disabling the DEC mechanisms (i.e., its long latency feedback loops for the gravity and the translation compensation, thus leaving only passive and short-latency proprioceptive stiffness). In the absence of the translation stimulus, noise alone produced no specific response. In contrast, noise injection in combination with the presentation of the translation stimulus led to a sway response that built up over repetitions. It showed features comparable to our experimental results with respect to the gain peak, the phase, and the coherence (**Figure 8A**). We take these results as further evidence for our resonance hypothesis.

## Vision-Induced Changes of the Sway Response

A straightforward hypothesis related to the above resonance concept is that the visual information, which we provided in the EO conditions, was damping the resonance. We dealt here with two visual signals (visual position information with EO/SI; visual position and velocity information with EO/CI), both of which we assume to antagonize in our experiments the resonance evoked by the translation stimulus (in terms of lowering the response peak in the upper MDR and allowing for sway in the LFR).

As detailed in our full SIP model (**Supplementary Appendix A1**), we attributed the reduction in sway response seen with EO/SI to a visual position signal that subjects used for compensation of the body lean disturbance. Accordingly, EO/CI then would add to the sensory lean estimate furthermore a visual velocity signal (presumed to stem foremost from expansion/contraction of visual flow fields during stimulus and body sway). In the form of a translational (tangential) head velocity signal, the same visual input was used in our model simulations after mathematical differentiation to improve the estimate of the support translational acceleration. Other sensory signals that are indispensable for the acquisition and functional integration of the visual information (e.g., current eye to head and head to body orientations) were here neglected for simplicity.

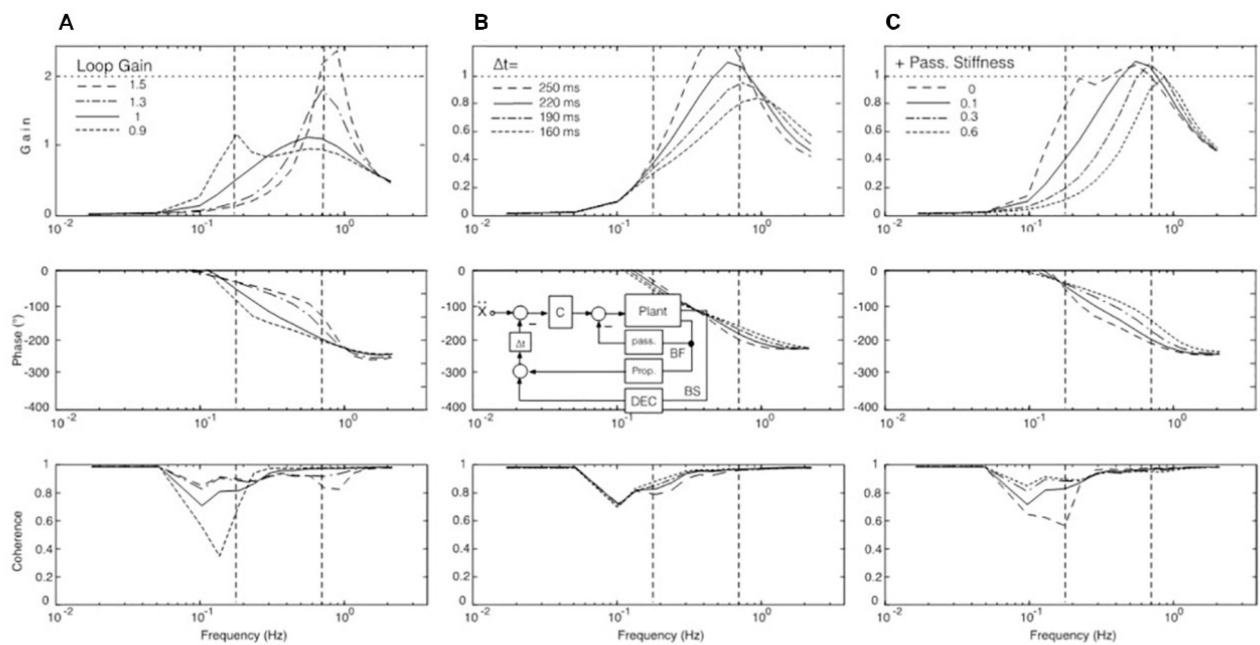

**FIGURE 7 |** SIP model simulations to test our hypothesis of a resonance effect. **(A)** Variations of loop gain. With a loop gain of unity, a peak occurs in the upper MFR (full line), but there is in addition also a tendency for a peak slightly below 0.2 Hz with reduced loop gain (dotted). Note also that the larger the peak is in the range of mid and high frequencies, the lower is the gain in the low-frequency range (by which we explain the absence of considerable sway in the low-frequency range with EC in our experiments; compare **Figures 3A–C**). Other parameter changes had different effects. For example, varying instead the time delay of the feedback loops increased, respectively, decreased the amplitude of the gain peak of the FRF (panel **B** with inset, simplified model), while increasing selectively the gain of the passive joint stiffness shifted primarily the gain peak toward higher frequencies **(C)**.

With the visual-only stimuli in our experiments, the small sway responses we obtained differed remarkably from the pronounced sway in the model simulation (**Figures 3D,E**). The reason, unmodeled in the simulation, obviously is that our subjects cognitively suppressed an immersion into an illusion of self-motion. The very small sway that was still obtained likely reflects enhanced insecurity (increased noise) of subjects experiencing their kinematic state, although they very likely "knew" that it was the cabin that was moving. These considerations led us to consider in our simulations (Figure 8B) a theoretical visual-only input signal (full lines, representing EO/CI input alone, which is insufficient for stabilization at low frequencies, so that the response develops here a very high gain).
The dashed curve in the figure gives the response with proprioceptive feedback alone (EC), and the dotted curve gives the results with the combination of both. The latter simulation resembles the experimental result obtained for sway in the EO/CI condition and supports our hypothesis that the visual input dampens the resonance tendency of the control loop.

Compared to these findings, there exists a seeming contrast to several studies reporting that postural sway can readily be evoked by visual field motion (see section "Introduction"; Lee and Lishman, 1975; Lestienne et al., 1977; Berthoz et al., 1979; Van Asten et al., 1988). These previous findings are, however, not necessarily in conflict with the present ones of an illusion suppression by cognition, if one accepts that the suppression can be overcome by a visual immersion effect given special stimulus conditions (like lasting exposures or low stimulus frequencies of the visual stimulus), which help to entrap subjects perceptually into a vection state (i.e., a compelling sense of self-motion; see Howard and Howard, 1994). In the present experiments, the fast and jerky motion of the stimulus likely helped to suppress the occurrence of vection in our subjects.

## Comparison to Previous Studies

As mentioned in Introduction, previous studies (Buchanan and Horak, 1999; Corna et al., 1999) reported for sinusoidal translation stimuli at the low frequencies of 0.1–1.25 Hz for eyes open a response pattern of "riding the platform," where legs and upper body moved approximately *en bloc* with the support surface. At higher frequencies, the upper body tended to sway less or remain almost stationary, while the lower body on the support below was swinging. With the eyes closed, the evoked sway changed and resembled generally more that of an inverted pendulum. Conceivably, using periodic stimuli might have allowed those subjects to involve prediction in their balancing. Yet, although our subjects reported on request the PRTS stimuli as unpredictable, there exist some similarities in the response patterns with the previous data in the sense that also our subjects showed with EC in the LFR a "riding the platform" pattern, while they were producing sway with a gain close to unity in the upper MFR and in the HFR (see EC responses in **Figure 3A**). We, therefore, conceive that biomechanical and physiological factors determined the previous responses more than prediction, and that previous and present data are, in principle, compatible.

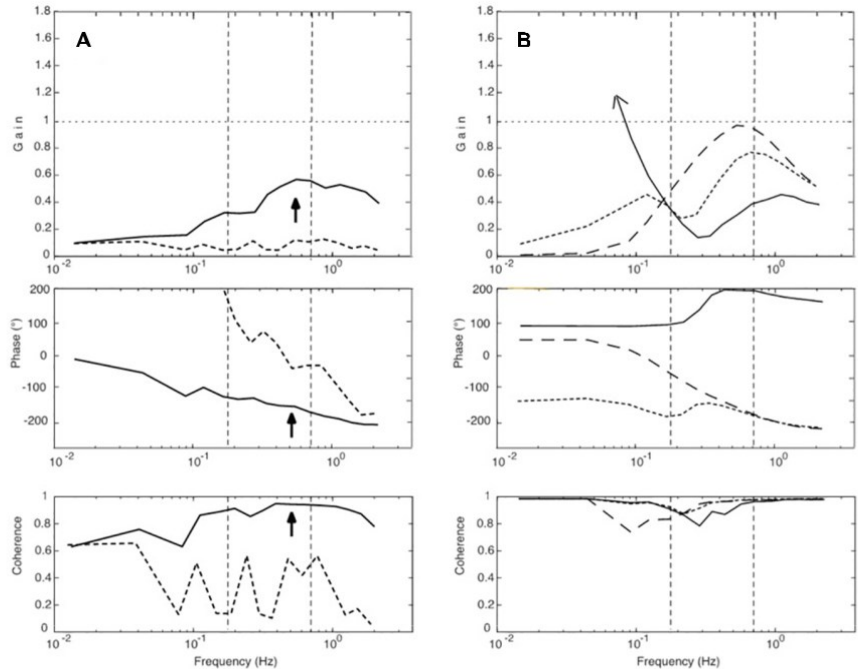

**FIGURE 8 | (A,B)** Further simulation results. **(A)** Effects of injecting low-frequency noise into the servo loop of the COM model (**Figure 4**). Dotted curves give simulation results obtained with noise alone. When adding a weak support translation stimulus, the response (full lines) developed a resonance peak and a coherence of almost unity in the upper MFR (arrow) with a phase lag of approximately 180 . **(B)** Results of simulations that tested the theoretical hypothesis of an antagonism between visual and proprioceptive effects. The dashed curves represent sway responses obtained with proprioceptive feedback alone, full lines those with continuous visual feedback alone, and the dotted curves give the results obtained with the combination of both (note its resemblance to **Figure 3C**).

### Additional Biomechanical Aspects

A remarkable feature in both, the experimental data of **Figures 3A–C** and the simulations of **Figures 5A–C** was that the gain at the highest frequency reached very similar endpoint values across the three visual conditions EC, EO/SI, and EO/CI. The similarity of these endpoints appears surprising because the gains in the low and mid-frequency ranges were affected by the illumination conditions to different degrees. We, therefore, hold that the endpoint region was determined mainly by the biomechanics of the system and explain the different endpoints seen in **Figures 3F,G** by a change in biomechanics (i.e., in terms of stiffening of ankle and hip joints).

The reason to performed the latter experiments with the en-bloc translation of both cabin and body support originally was to dissociate between the visual effects of the two disturbance estimators in the DEC model, i.e., one for the support surface translation and another for the evoked body lean and gravity effect (compare model in **Figure 4**). However, the simulation results for both experiments resembled each other. Closer inspection showed that the response was with both stimuli in fact dominated by the body lean estimator (i.e., the gravity compensation), while the effects of the vestibular and the visual linear acceleration signals from the linear acceleration estimator were weak – which we attribute to the fact that they, as second derivative terms in the feedback loop, would otherwise be prone to control instability (Mergner, 2010).

As underlined by Jilk et al. (2014), the postural control of the COM is functionally more relevant than the control of single joints for the balancing, which is in line with several earlier studies (e.g., Krishnamoorthy et al., 2005; Hsu et al., 2007). This relates to the often emphasized concept of the ankle and hip strategy (Horak and Nashner, 1986; Kuo and Zajac, 1993; Gatev et al., 1999; Runge et al., 1999). With moderate perturbations, the default is the ankle strategy. With increasing disturbance magnitude and balancing difficulty, it becomes accompanied and eventually replaced by the hip strategy. Behind this, not always emphasized enough, is the main objective of postural control, which is to maintain the body COM over the base of support (Massion, 1992) and the selection of the best method to realize this. For example, the absence of a given sensory input such as vision may facilitate the shift from one to another strategy.

In face of the complexity in which the COM stabilization is to be achieved, it has been pointed out that neither a SIP nor a DIP model alone is a good model to describe how the COM is stabilized (Hsu et al., 2007). Yet, COM control is critical for balancing and, in fact, may be achieved in different body configurations (Jilk et al., 2014). Therefore, we hold that considering a SIP model, alone or extended into a DIP model, can elucidate specific important aspects of the control.

In the study of Lippi et al. (2020), the same translation stimulus as used here evoked considerable excursions of the HAT segment in the hip joints, this in addition to the body COM excursions in the ankle joints. To appreciate corresponding effects for the balancing in the present context, we consider in **Supplementary Appendix B** simulations performed with a DIP model (compare **Figure 4**) as well as biomechanical calculations. We show that the trunk sway on top of the hips exerts, in addition to the damping by the visual velocity and position cues from the stationary visual surroundings, a *mechanical damping* of the excursion of the body COM in space. Interestingly, the study of Lippi et al. (2020) showed that the presence of visual orientation cues considerably reduced also the sway of the HAT segment around the hip joints, so that this indirect route via a visual effect on trunk sway in the hip joints appears to be relevant for the COM stabilization as well.

## Conclusion

This study suggests that humans, when balancing support surface translations in the earth-horizontal plane, stabilize upright body posture primarily on the basis of proprioceptive feedback control mechanisms in the ankle and the hip joints. Per se, this control is, due to time delays in the sensorimotor feedback control, prone to resonance effects, which tends to enhance the disturbance-related body lean effects. Our study indicates that vision as well as biomechanical effects arising with hip bending reduce the resonance tendency and thus help the postural body stabilization.

## DATA AVAILABILITY STATEMENT

The raw data supporting the conclusions of this article will be made available by the authors, without undue reservation.

## ETHICS STATEMENT

The studies involving human participants were reviewed and approved by Ethics Committee of the Freiburg Clinics. The patients/participants provided their written informed consent to participate in this study.

## AUTHOR CONTRIBUTIONS

All authors listed have made a substantial, direct, and intellectual contribution to the work and approved it for publication.

## FUNDING

This research was funded by International Post Doc Research fellowship to EA from the Scientific and Research Council of Turkey (Tubitak), Scientific Human Resources Development, and the Eurobench Project (EU FP7 Grants 600698 and 610454).

## SUPPLEMENTARY MATERIAL

The Supplementary Material for this article can be found online at: https://www.frontiersin.org/articles/10.3389/fnhum.2021.615200/full#supplementary-material